\documentstyle[12pt]{article}

\catcode`@=11
\@addtoreset{equation}{section}
\catcode`@=12

\newcommand{\be}{\begin{equation}}
\newcommand{\ee}{\end{equation}}
\newcommand{\sst}{\scriptscriptstyle}

\newcommand{\bea}{\begin{eqnarray}} \newcommand{\eea}{\end{eqnarray}}
\def\r0{r_{\sst 0}}

\begin{document}

\title {\bf Covariant treatment of flavour oscillations}

\author{\bf Subhendra Mohanty\\
 \normalsize Physical Research Laboratory,
Navrangpura, Ahmedabad 
380009,
INDIA\\}
\maketitle
                                                          
\begin{abstract}
\normalsize\noindent

{
We perform a covariant wavepacket analysis of neutrino oscillations 
taking
into account the  lifetime of the neutrino production process . 
We find that flavour oscillations in space are
washed out when  the neutrinos are produced from long lived resonances
 - and what may be observed in appearance/disappearance experiments is a
uniform
conversion probability
independent of distance.
 The lifetime of the resonance which produces the
neutrinos acts as the the effective baseline of the experiment.
 For this reason  the LSND experiment where
neutrinos
are produced from muon decay   has two orders
of magnitude more sensitivity to neutrino mass square difference than
other experiments where the neutrinos are
 produced from pion or kaon decays.Wea also show that there are no EPR
type oscillations of the secondary decay particles.
} 
\end{abstract}
\bigskip\bigskip

\section{Introduction}

There are two different formulas which describe flavour oscillations in
spacetime; the  one applied to  neutrino oscillations \cite{books} 
is derived in the high
energy regime 
and the other  applied to  kaon oscillations \cite{kaons}  is valid
in the non-relativistic
limit. 
From the derivations of these formulas it is not clear what formula is
 applicable at intermediate energies (for example for kaon oscillations
from stopped protons where $(m^2/E^2)\sim 25\%$) . 
A covariant
derivation of a flavour oscillation
formula which would be valid at both high and low energies 
would be of interest from the conceptual as well as the experimental point of
view.  

  Kayser and Stodolsky \cite{KS} have  advocated a                
covariant
generalisation of the non-relativistic phase factor $exp (-imt)$ by
replacing the absolute time $t$ by the Lorenz invariant proper time
$s=(t^2 -x^2)^{1/2}$.  In the lab frame the covariant expression for the 
phase 
factor  is $exp -im_i s_i= exp ~(-i m_i^2 t / E_i)$. The phase difference
between
two mass eigenstates is $\Delta (m^2 /E) \simeq (\Delta m^2 /E)$ which is
twice the the phase difference of the
standard formula \cite{books}. 
This result has prompted the claim  \cite{SWS}  that
 kaon oscillations in $\Phi$ factory will have oscillation length which
is half of what is given by the standard formula. This claim has been
refuted by   
\cite{gold} who take the view that the interference
phase difference should not be evaluated at different space-time points
but should be evaluated at the average spacetime interval and taking
the
phase difference to be $(m_1-m_2) (s_1 +s_2 )/2$ instead of $(m_1 s_1 -m_2
s_2)$ they recover the
standard
formula.
Other methods of
showing that an extra factor of two does not appear in the kaon
oscillations formula have been discussed in \cite{lip2}.
A covariant derivation of the neutrino oscillation formula has been given
by Grimus and Stockinger \cite{grimus} who treat the entire process of
neutrino production ,propagation and detection as a single Feynman
diagram. They show that on taking the large distance limit of the neutrino
propagator the scattering cross section shows a space-time
oscillatory behaviour and the oscillation length is identical to that 
given by the standard formula \cite{books}. G-S assume
the initial states to be plane waves therefore the concept of a coherence
length \cite{coh} does not emerge in their formulation.

The oscillation phenomenon occurs because particles are produced and
detected as weak interaction eigenstates (the neutrino states $\nu_e,
\nu_\mu$ and $\nu_\tau$ or the Kaon states $K^0 , \bar K^0$ etc)
but the propagators are diagonal in the mass eigenstates
(the neutrino mass eigenstates $\nu_i, i=1-3$ or the Kaon mass
eigenstates
$K_L , K_S$). The probability
amplitude for oscillations
of a gauge eigenstate ($|\alpha>$) to another ($|\beta>$) is a lof a gauge
eigenstate ($|\alpha>$) to another ($|\beta>$) is a linear
superposition
of the
propagation amplitude of the mass eigenstates ($|i>$).
\be
{\cal A}(\alpha\rightarrow \beta;t)= \sum_i~ <\beta|i>~ 
<i|e^{-i Ht}|i><i|\alpha>
\label{ampl} 
\ee
In relativistic field theory the propagation amplitude of the maIn
relativistic field theory the propagation amplitude of the mass
eigenstates $<i|e^{-i Ht}|i>$
 can be identified with the Feynman propagator. At large time-like
spacetime separation, the Feynman propagator (for both fermions and
bosons) has the form,
\be
K(m_i, t)\simeq~~~~
({m_i\over2\pi i s_i })^{3/2} ~e^{-i m_i s_i}
\label{sf1}
\ee
where $s_i$ is the invariant space-time interval
propagated by the $\nu_i$ mass eigenstate. In the lab frame the phase
factor of the amplitude (\ref{sf1}) is given by
$exp ~\left({-i m_i^2 t \over E_i} \right)$.In this paper we derive the
oscillation amplitude by evaluating the
Feynman propagator of the neutrinos in the large time-like asymptotic
limit. The asymptotic propagator of a position eigenstate has the form
$K (x,t; m_i) \simeq ( m_i / 2\pi i  s_i )^{3/2} ~exp{-i m_i s_i}$ where
$s_i$ is the spacetime interval propagated by the $m_i$ mass eigenstate.
We derive the same expression in the path integral method. Using the path
integral method we show that for large distance propagation the forward in
time paths are equivalent to on-shell particles.

Another aspect of flavour oscillations on which there is no consensus of
opinion is
the question of the energy momentum of a flavour eigenstate. A flavour
eigenstate propagates as a linear combination of different mass
eigenstates which are on-shell. Some authors  have argued that the
different mass eigenstates have
common energy  but different mommenta $P_i= \sqrt{ E^2 - m_i^2}$ while 
others have assumed that
the states have same mommenta but different energy 
$E_i= \sqrt{P^2 + m_i^2}$.
We show here that different mass eigenstates in a linear combination of
states have niether energy nor mommentum in common. For long
distance propagation, the energy and  mommenta of a given mass eigenstate 
are not independent however but are
related by the mass shell condition. 
 We use these results to
show that there are no EPR type associated particle oscillations as has
been claimed in literature \cite{SWS}. This result is in agreement with
the conclusions of Lowe et al. \cite{gold}
and Dolgov et al. \cite{lip2} . Our proof however is more direct in that
we show that although the energy and momenta of the muons depend on the
outgoing  neutrino mass, the phase difference of the muon states
corressponding to different neutrino masses cancels exactly and there is
no spacetime dependence of the muon probability distribution.

 If the initial wavefunctions  of the propagating particles were 
 strictly delta functions in spacetime then no interference between
different mass eigenstates can take place. We therefore
generalise the delta function propagators to propagators of gaussian
wavepackets. The interference term as a function of distance
 is obtained by taking the time-overlap
of different mass eigenstate propagators . The
expression for flavour conversion probability (for say two flavours with
mixing angle $\theta$) as a function of distance $X$ turns out to be,
\be 
P(\alpha  \rightarrow \beta; X)= {1\over 2} sin^2 2\theta 
(1- cos({\Delta m^2 \over 2 P} X) ~e^{-A})
\label{osc1}
\ee
Therefore extra factor of two which came from naively subtracting
the
phases of 
plane wave propagators \cite{SWS} goes away in the wavepacket averaging
and the standard expression for the oscillations length is recovered. This
is in agreement with \cite{gold}-\cite{lip2}. The covariant wavepacket
treatment
introduces a new contribution to the exponential factor $A$ when particles
are produced from long lived resonances ( for example for neutrinos from
muon decay as opposed to  neutrinos from Z decay). At distances smaller
than the
coherence length , the exponential suppression factor is,
\be
A~~=~~ ( {\Delta m^2 \over 2 \sqrt{2} E}~ \tau)^2
\label{Ab}
\ee 
where $\tau$ is the lifetime of the resonance which produces the
particle which undergo flavour oscillations. If 
the uncertainty in position of the
initial
particle $v \tau$  is larger than
the detector distance $X$ , the exponential
term washes out the oscillations in (\ref{osc1}). 
In  neutrino experiments
where the source is pions, kaons, muons or nuclear fission,  
the spatial oscillations of the conversion probability cannot be observed.
What can be observed in these experiments  is a constant (distance
independent) conversion
probability. 
  The conversion probability  is sensitive to 
values of $\Delta m^2 \simeq (2\sqrt 2
E/\tau)$ , which means that the experimental bound on $\Delta m^2$ is
lower with longer
-lifetime sources. For this reason the
LSND
experiment \cite{LSND} which uses neutrinos from muon decay ($\tau_{\mu} =
2.19 \times 10^{-6}s$) is sensitive to two orders of magnitude lower
neutrino mass
square difference compared
to other accelerator experiments like BNL-E776 \cite{E776}, Karmen
\cite{karm} and CCFR \cite{CCFR} which use neutrinos from pion and kaon
decays ($\tau \sim 10^{-8} sec$). We fit the experimental data from LSND
along with BNL-E776, Karmen, CCFR and Bugey \cite{BUGEY} experiments with the
conversion probability formula (\ref{osc1}) and plot the allowed range
for the mass difference and mixing angle. We find that there is a large
region of LSND
allowed band which is not ruled out by the other experiments. 
The region of $\Delta m^2$ and $sin^2 2\theta$ which 
is allowed by LSND and all other experiments is $2.2 \times 10^{-3} eV^2
 < \Delta m^2 < 6.3 eV^2 $ and  $0.002 <   sin^2 2 \theta < 0.15$. The
lower bound on allowed $\Delta m^2$ is two orders lower than what is
obtained by fitting the LSND data with the standard oscillation formula
(equation (\ref{osc1}) with $A$ set to zero).

\section{Field theoretic derivation of phase factor}
In relativistic field theory the propagation amplitude of the mass
eigenstates $<i|e^{-i Ht}|i>$
 can be identified with the Feynman propagator
 $S_F(x_f-x_i,m_i)= <T~\nu_i(x_i)
\bar \nu_i(x_f)>$. The Feynman propagator in position space is , 
\be
K(x,m_i) = -i \int {d^4 p \over (2\pi)^4} {(i \not \partial +
m_i)\over 
p^2 - m_i^2 -i \epsilon} e^{-i p\cdot x} .
\label{two}
\ee
This integration  can be done by  expressing the
denominator  as an exponential,
\be
{-i \over p^2 - m_i^2 -i \epsilon} = \int_0^{\infty} 
d\alpha~~
exp~ \{i \alpha ( p^2 - m_i^2 -i\epsilon)\}
\label{twoa} 
\ee
 and integrating
over the resulting Gaussian in $p$.
The remaining integral over $\alpha$,
\be
\int_0^{\infty} d \alpha ~\alpha^{-2} ~exp~ \{i\alpha m_i^2 + i{(t^2-
|{\bf x}|^2)\over 4 \alpha}\}
\label{twod}
\ee
can be performed by substituting $s=(t^2- |{\bf x}|^2)^{1/2}$, $z=i m s$
and
$\eta =2(\alpha m /s)$ and making use of the integral formula
\cite{grad} for the 
Bessel function    
\be
{1\over 2} \int_0^{\infty} d \eta ~ \eta^{-(\nu +1)}  ~exp ~
-{z\over2}(\eta+{1\over \eta}) = K_\nu (z).
\label{twoe}
\ee
The resulting expression for the Feynman propagator (\ref{two}) is
\be
K(x,m_i) = ({i\over 4 \pi^2}) (i \not \partial + m_i )~({m_i \over
s_i})~ K_1(i m_i s_i) 
\label{three}
\ee
where  $s_i= (t^2 - {\bf x}_i^2 )^{1/2}$ is the
invariant spacetime interval propagated by the $\nu_i$ mass eigenstate. If
this interval is large ($s >> m_i ^{-1}$) and time-like ( $t \ge |{\bf
x}|$ ) then we can use the asymptotic expansion of the Bessel
function 
\be
K_1(i m s) \simeq \left({2 \over \pi i m s} \right)^{1/2} ~e^{-i m s}
\label{bessel}
\ee
to obtain from (\ref{three}) the expression for the propagation amplitude
at large time-like separation
\be
K (x,t; m_i) =\left({m\over 2\pi i
\sqrt{(t_f-t)^2 -({\bf x_f} - {\bf x_i})^2)}}\right)^{3/2} ~ exp \{-im
\sqrt{(t_f-t)^2
 -({\bf x_f} - {\bf x_i}^2)} \}
\label{four} 
\ee
The phase factor of the amplitude (\ref{four}) is Lorentz-invariant and
can be
written  in terms of the neutrino energy $E$  and the time of
flight
$t$ as measured from the lab frame as
$-im_i s_i =  -i m_i~ (1-v_i ^2)^{1/2} ~t =  -i (m_i^2~ / E_i)~t$.

An
extra factor of two appears on 
subtracting the phases at different
spacetime points. A flavour eigenstates neutrino or kaon is
observed at a single spacetime point and one should therefore compute the 
phase difference at the overlap of the two mass eigenstate wavepackets.
This averaging over spacetime is done formally  by considering the
propagators of gaussian wavefunctions
as opposed to plane waves , and the standard expression for the
oscillation length is recovered.

\section{ Conservation laws for long distance propagators}

A flavour
eigenstate propagates as a linear combination of different mass
eigenstates which are on-shell. 
The phase difference is obtained by some authors by assuming that
 the different mass eigenstates have common energy, and by some by
assuming
that they have a common momentum \cite{lip}. 
 In this
section we show that the linear combination of different mass eigenstates
have neither the same
energy nor
the same momentum. We show that 
 particles propagating  over large distances $(X >> P/m^2) $ are
constrained by the 
conservation laws at the vertex to be on shell. The on-shell condition is
all that is needed to fix the phase difference and the oscillation length.
 
Consider a diagram with a  propagator between vertices at spacetime
points $(x_1, x_2)$ with a number of external
legs at the vertices. The amplitude is proportional to
\be 
\int d^4x_1~~d^4x_2 ~~exp(-i \sum_i q_i \cdot x_1 ) ~~K(x_1,x_2)~~exp(
i\sum_f q_f \cdot x_2)
\label{c1}
\ee
Where $q_i$ are the incoming four momenta from the external legs at $x_1$
and $q_f$ are the
incoming four mommenta from the external legs at $x_2$.
Substituting the  propagator
\be
K(x_1,x_2)= \int d^4 p {e^{-i p\cdot (x_1-x_2)}\over p^2 -m^2 +i \epsilon}
\label{c2}
\ee
in (\ref{c1}) and integrating over $x_1$ and $x_2$ gives the usual energy
mommentum conservation $\delta^4 (\sum_i q_i - p)$ and $\delta^4
(\sum_f q_f - p)$ at the vertices. When the proper distance between the
two vertices $s$ is larger than $m^{-1}$, then the  propagator used in
(\ref{c1}) is
the  asymptotic form (\ref{four}), which in mommentum space can
be written
as \cite{bogol}, 
\bea
K(x_1,x_2; s>> m^{-1})&=& {1\over (2 m s)^{3/2}} e^{-i m
s}\nonumber\\[8 pt]
&=&\int {d^4 p \over \sqrt{{\bf P^2} +m^2}} ~~exp~\{-i(\sqrt{{\bf P^2}
+m^2}})(t_f-t_i)
 +i {\bf P}\cdot{\bf(x_f-x_i)\}
\label{c3} 
\eea
Substituting (\ref{c3}) in (\ref{c1}) and integrating over $x_1$ we obtain
\bea
\int d^4 x_1~ exp~ \{-i  (\sum_i E_i - \sqrt{{\bf P^2} +m^2}~~)t_1 +
i (\sum_i
{\bf q_i}- {\bf P})\cdot {\bf x_1}\} \nonumber\\[8pt]
=~\delta (\sum_i E_i - \sqrt{{\bf P^2}
+m^2})~~\delta^3(\sum_i
{\bf q_i}- {\bf P}) 
\label{c4}
\eea
This implies that at a vertex
energy and
mommentum are  conserved and
in addition those particles which propagate without freely over distances
larger than $(|{\bf P}|/m^2) $ obey the mass shell constraint ,
$E= \sqrt{{\bf P}^2 + m^2}$.

 \section{Path Integral method}

From the derivation of the propagation amplitude  (\ref{four})
it is clear that there is no difference in the result on whether the
propagating fields are bosons or fermions. The same result for the
propagation  amplitude is therefore also expected in 
relativistic quantum mechanics . 
In relativistic quantum mechanics the propagation amplitude $K(m_i,t)$ 
is computed from the classical action for a relativistic particle
by the path integral method \cite{feyn}. The amplitude for a particle of
mass $m_i$ to
propagate from the spacetime point $x_i$ to another spacetime point $x_f$
can be written formally as,
\be
K(x_f,x_i;m)\equiv \int {\cal D}[x(\tau)]~~ exp ~\{
-{i m\over 2} \int_{x_i}^{x_f}  \left(
\eta_{\mu \nu}
{dx^{\mu}\over d\tau} {dx^{\nu}\over d\tau}+1 \right)d\tau \}
\label{pi1}
\ee

For infinitesimal propagation between points $x_j$ and $x_{j+1}$ the
quantum mechanical
transition amplitude may be written as
\be
K(x_{j+1} , x_j) = C~exp\{-{i m \over 2}
\left( {(x_{j+1} - x_j)^2
\over (\Delta \tau)^2} +1 \right)\Delta \tau \}
\label{ki}
\ee
The condition 
$K(x_{j+1} , x_j)= \delta^4(x_{j+1}-x_j)$ as $\Delta \tau \rightarrow 0$
 fixes the normalisation constant  $C=\left( m / 2 \pi i \Delta \tau\right)^2$.
In what follows we will not write out the explicit form for the
normalisation factors as
the final normalisation can be fixed from the boundary
conditions.

To carry out the path integral we divide each path into $N\rightarrow
\infty$ infinitesimal segments of length
$\Delta \tau \rightarrow 0$ such
that $N \Delta \tau = \tau $ the total path length. The amplitude for
propagation from $x_i$ to $x_f$ is expressed as a product of the
infinitesimal amplitudes $K(x_j, x_{j-1})$ over the small straight 
line segments $\Delta
\tau$,
\be
K(x_f,x_i)= \int dx_1 \cdots dx_{N-1}~~ K(x_f, x_{N-1}) \cdots K(x_1,
x_i)
\label{pathint}
\ee
By substituting the infinitesimal amplitudes (\ref{ki}) in the expression
(\ref{pathint}) we have  the amplitude $K(x_f,x_i)$ as a product of ordinary 
integrals of the
infinitesimal amplitudes $K(x_j, x_{j-1})$   
over the intermediate spacetime points $(x_1, \cdots x_{N-1})$.  
\be
K(x_f,x_i)= C^{\prime}~~ \int dx_1 \cdots dx_{N-1}~ ~exp~\{ {-im\over 2
\Delta
\tau} \left(\sum_{j=1}^{N} \left(x_j-x_{j+1}\right)^2  + N\Delta
\tau \right) \}
\label{k}
\ee
 The
Gaussian integrals in (\ref{k}) can be performed sequentially 
and the result of  the $N-1$
 integrations is
\be
\int dx_1 \cdots dx_{N-1}~ ~exp~\{ {-im\over 2 \Delta
\tau} \left(\sum_{j=1}^{N} \left(x_j-x_{j+1}\right)^2 \right) \}
 = C_{N}~~ exp \{\left({-im\over 2\Delta \tau}
\right) \left( {1\over N }(x_i - x_f)^2 \right)\}
\label{gn}
\ee
 Using (\ref{gn}) in  (\ref{k}) and writing $N \Delta \tau = \tau$ in the
limit $N\rightarrow \infty$ and $\Delta \tau \rightarrow 0$ , we have the
expression for the propagation amplitude $K(x_i, x_f)$ as a function of
the proper time of $\tau$ along some path,
\be
K(x_i,x_f,\tau) = \left({m\over 2\pi \tau}\right)^2 ~ exp \{{-im \over 2}
\left({(x_i - x_f)^2 \over \tau } + \tau \right)\}
\label{ktau}
\ee
where the normalisation constant has been fixed by the requirement that
$K(x_i,x_f,\tau) = \delta^4(x_i - x_f) $ as $\tau \rightarrow 0$.
    
In relativistic quantum mechanics the proper time along a path $\tau$ is
an
extra parameter that must be integrated over to include all paths
including those which go backwards in the coordinate time $t$. The
relativistic propagation amplitude from (\ref{ktau}) is
\be 
K(x_f,x_i)=  \int_{0}^{\infty} d\tau 
 \left({m\over 2\pi \tau}\right)^2 ~ exp {-im \over 2}
\left({(x_i - x_f)^2 \over \tau } + \tau \right)
\label{krel}
\ee
The integral over $\tau$ can be performed by using the formula
(\ref{twoe})to give the amplitude (\ref{krel}) as
a Bessel function 
\be
K(x_f,x_i)= \left(m\over 2 \pi i\right)^2 {2\over s} ~K_1(i m s)
\label{kf}
\ee
which is of the same form as the field theory propagator(\ref{three})its  
asymptotic
form 
\be
K(x_f,x_i;s>>m^{-1})    \simeq \left({m\over 2 \pi s}\right)^{3/2}
exp (-i m s)
\label{ks}
\ee
is the same as (\ref{four}).

The expression given in (\ref{kf}) is the correct form of the propagator
for particles which in general are not on-shell. To
compute the propagator for strictly on-shell (real) particles, we
 and restrict ourselves  only
to those paths which go forward in time and propagate over a fixed total
proper time
$\tau=s=\sqrt{(x_f-x_i)^2} $. The on-shell propagator is then obtained by
inserting $(2 \pi \tau/m)^{1/2}~\delta(\tau-\sqrt{(x_f-x_i)^2})$ (the
factor of $(2 \pi \tau/m)^{1/2}$ is for normalisation) in (\ref{krel}) to
give, 
\be
K(x_f,x_i)_{on-shell} \simeq \left({m\over 2 \pi i s}\right)^{3/2}
exp (-i m s)
\label{kos}
\ee
Comparing (\ref{kos}) and (\ref{ks}) we see that in the 
asymptotic limit the
off-shell propagator becomes identical  to the on-shell propagator. In
other words `virtual' particles become `real' on
large distance propagation.

In non-relativistic quantum mechanics the proper time
$\tau=(t_f-t_i)$ the coordinate time interval. The non-relativistic
propagator (which off-course is always on-shell) is obtained by inserting 
$(2 \pi \tau/m)^{1/2}~\delta(\tau-(t_f-t_i)^2)$ in (\ref{krel}). The
non-relativistic propagator thus obtained 
is the familiar form
\cite{feyn},
\be
K_{NR}({\bf x_i},t_i;{\bf x_f},t_f) = \left({m\over 2\pi i
(t_f-t_i)}\right)^{3/2} ~ exp \{-im (t_f - t_i)\left(1 -{({\bf x_f} - {\bf 
x_i})^2
\over
2 (t_f-t_i)^2 } \right)\}
\label{knr}
\ee
The amplitude (\ref{knr}) has the same form as the non-relativistic limit
of the asymptotic propagator (\ref{four}). As discussed earlier the phase
factor $exp -imt$ implies that in 
interference between two different mass
states of non-relativistic particles (like $K_L$ and $K_S$) the
oscillations are periodic in $(\Delta m) t$.

\section{Wavepacket analysis}

The propagator $K(x_f, x_i)$ given in (\ref{four} ) is the amplitude for
the
propagation of a
particle localised at $x_i$ (a delta function initial wave function) to
be detected at $x_f$. The propagator of some general initial 
 wavepacket $~\Psi_{in} (x-x_i)$ is obtained by
using  the expressions for the delta function propagator given in
(\ref{four}) and the
superposition principle,
\be
\tilde K(x_f-x_i)= \int d^4x ~K(x_f,x) ~\Psi_{in} (x-x_i)
\label{psif}
\ee
If 
the
initial wave function is a gaussian ,
\bea
\Psi_{in}(x-x_i)=N~ exp \{-i (E_a (t-t_i) +i {\bf P_a}\cdot ({\bf
x-x_i})\}~~
exp ~~\{ -{({\bf x-x_i})^2 \over 4
\sigma_x^2} -{(t-t_i)^2 \over 4
\sigma_t^2}  \}
\label{psii}
\eea
where $\sigma_x$ and $\sigma_t$ are the uncertainties is the initial
position and time of production of the particle respectively and $N$ is
the normalisation constant.
In the
earlier wave-packet analyses of oscillation problem \cite{coh}, the
uncertainty
in the time of production was neglected. In a covariant treatment both
should be included. In the next section we show that  the time uncertainty
gives rise to a novel phenomenon of conversion without oscillations in
many experimentally relevant situations.
Substituting the expression for $K(x_f,x)$ given in
(\ref{four}) and the gaussian initial wavepacket (\ref{psii}) in
(\ref{psif}) and evaluating the 
 integral by the stationary phase method
\cite{wolf} we obtain the expression for the  propagation amplitude
of a gaussian wavepacket ,
\be
\tilde K({\bf X},T;m_a)=({1\over 4 \pi |{\bf X}|^2})^{1/2}(
{\pi\over
\sigma_x^2 + v_a ^2
\sigma_t^2})^{-1/4}  exp\{-i{m_a^2 \over E_a} T
+i {\bf P_a}\cdot ({\bf X} -{\bf v_a}T ) \} 
~exp\{ - {({\bf X} -{\bf v_a}~T  )^2\over 4( \sigma_x^2 + v_a ^2
\sigma_t^2)}\}
\label{wp}
\ee
where ${\bf X}= {\bf x_f - x_i}$ and $T=t_f-t_i$ are the space and time
intervals
propagated
by the center of the gaussian wavepacket and ${\bf v_a}\equiv{\bf
P_a}/E_a$ . The normalisation constant has been fixed such that
$\int~ 
dT~ d^3 {\bf X}~ \tilde K({\bf X},T; m_a)~~ \tilde K^{\dagger} ({\bf
X},T;m_b) = 1$.
The factor of $(1/4 \pi |{\bf X}|^2)^{1/2}$ in  (\ref{wp}) shows that the
particle flux
decreases inversely with the sqaure of distance.
Neutrino disappearance experiments  look for
evidence of depletion of a certain neutrino species over and above the
expected inverse square decrease in 
flux. In the following we will not display this factor
 $(1/ 4 \pi |{\bf X}|^2)$ in the probability expressions.
The amplitude for the oscillation of
one type of neutrino flavour to  another 
is obtained by substituting the propagation amplitudes for mass
eigenstates (\ref{wp}) in ,
\be
{\cal A}(\alpha\rightarrow \beta; X,T) = \sum_{a}^{3} U_{\alpha
a}~ \tilde K(m_a;X,T)~~~U^*_{a\beta} 
\label{Amp}
\ee
where $\alpha, \beta $ denote the flavor eigenstates ($\nu_e, \nu_\mu,
\nu_\tau$
or $K^0 ,\bar K^0$) and the summation index $a$ denotes a mass eigenstates
($\nu_1,
\nu_2, \nu_3$ or $K_L,K_S$) . $U_{\alpha a}=<\alpha|a> $ and
$U^*_{\beta a}=
<a|\beta> $ are the
elements of
the mixing matrix which relate the flavor eigenstates with the mass
eigenstates.
The probabilty of flavour oscilation as a function of space-time is
the modulus squared of the amplitude (\ref{Ab}) 
\bea
P(\alpha \rightarrow \beta;X,T)&=&|\sum_{a}^{3} U_{\alpha
a}~  \tilde K(m_a;X,T)~~~U^*_{a
\beta} |^2 \nonumber\\[8pt]
= \sum_a |U_{\beta a}|^2~|U_{\alpha a}|^2~~&+&~~\sum_{a\not=b}~U_{\beta
a}~U^*_{\alpha
a}
U^*_{\beta b}~U_{\alpha b}~~\tilde K({\bf X},T; m_a)~~ \tilde K^{\dagger} ({\bf
X},T;m_b)
\label{PXT}
\eea

 In interference experiments the time of flight of the
particle is not measured and  only the
distance between the source
and
the detector is accurately known \cite{lip} . The probability for the flavour
conversion as a function of distance 
 is given by the time integral of (\ref{PXT}) ,
\bea
P(\alpha \rightarrow \beta;X)&=&\int~dT~|\sum_{a}^{3} U_{\alpha
a}~  \tilde K(m_a;X,T)~~~U^*_{a\beta} |^2 \nonumber\\[8pt]
= \sum_a |U_{\beta a}|^2~|U_{\alpha a}|^2~~&+&~~\sum_{a\not=b}~U_{\beta
a}~U^*_{\alpha
a}
U^*_{\beta b}~U_{\alpha b}~~\int~ 
dT \tilde K({\bf X},T; m_a)~~ \tilde K^{\dagger} ({\bf
X},T;m_b)
\label{PX}
\eea
The interference term  given by the
time-overlap of the propagation amplitudes (wave-functions) of different
mass eigenstates can be evaluated for the gaussian propagator (\ref{wp})
and is given by,
\bea
 Re ~\int dT~~\tilde K({\bf X},T; m_a)~~ \tilde K^{\dagger} ({\bf
X},T;m_b)
&=& ~
~{2\over v}~ cos
\{ ({\bf P_a -P_b})\cdot {\bf X} - (E_a -E_b){({\bf v_a +
v_b})\over 2 v^2}\cdot {\bf X} \}\nonumber\\[8pt]
~ \times exp&\{-&(E_a-E_b)^2 {( \sigma_x^2 +
 v^2\sigma_t^2)\over 2  {v^2}}- {({\bf v_a - v_b})^2 ~X^2
\over 8
 ( \sigma_x^2 +   {v^2} \sigma_t^2)   {v^2}} \}    
\label{intX}
\eea
where  $  {v}\equiv \sqrt{(v_a^2 + v_b^2)/2}$.
 In terms of the average momentum ${\bf P}$ and
energy $E$
and their respective differences $\Delta {\bf P}$and
$\Delta E$ the interference term (\ref{intX}) ,to the
leading order in
$(\Delta P/ P)$ and $(\Delta E/ E)$, reduces to the form,
\be
~{2\over v}~ cos
\{{X\over  P}( E \Delta E - {\bf P} \cdot \Delta {\bf
P})\}~exp \{-A\} 
\label{I}
\ee
where the exponetial damping factor,
\be
A=(\Delta E)^2~ 2 \bar \sigma^2~({ E^2\over
P^2})+ ({\Delta P\over P})^2
~{X^2 \over 4  \bar \sigma^2}    
\label{AI}
\ee
and $\bar \sigma^2 \equiv ( \sigma_x^2 + (P/E)^2 \sigma_t^2)$.

In general neither $\Delta E$ nor 
$\Delta P$ is zero and they depend upon how the state is prepared.
For example if the mommentum of the initial and the associated final state
is measured then $\Delta P=0$. 
 In the last section we have shown
that for long distance propagation ($s > m^{-1}$)
 ${\bf P}$ and $E$  are not
independent and are related by the mass
shell constraint.
The particular combination that appears in the phase difference
 turns out to be independent of the
preparation and is fixed by the condition that each of the mass
eigenstates be on shell,
\bea
{E_i}^2 -{P_i}^2 &=& {m_i}^2   ~~~~~~~~i= a,b \nonumber\\[8pt]
\Rightarrow~~ 2 E \Delta E -2 {\bf P}\cdot  {\bf \Delta P} &=& \Delta m^2
\label{mshell}
\eea
Substituting (\ref{mshell} ) in  (\ref{I}) we see that the
interference 
term of  (\ref{intX}) is given by ,
\be 
{2\over v}~cos ( {\Delta m^2 \over 2 P} X) ~~e^{-A}~~~~~~~~~~,  
\label{cosA}
\ee
The probability for the flavour
conversion as a function of distance (\ref{PX}) is therefore,
\bea
P(\alpha \rightarrow \beta;X)&=&
 \sum_a~{1\over v_a}~ |U_{\beta a}|^2~|U_{\alpha
a}|^2~~\nonumber\\[8pt]
&+&~~\sum_{a\not=b}~ {1\over v}~~|U_{\beta
a}~U^*_{\alpha
a}
U^*_{\beta b}~U_{\alpha b} | cos( {(m_a^2 -m_b^2)\over 2  P}X
~~+~~\delta)~~e^{-A}
\label{covP}
\eea
where  $~\delta
=arg(U_{\beta a}~U^*_{\alpha a}
U^*_{\beta b}~U_{\alpha b} )$.

  We see that the
 standard oscillation formula which was obtained for relativistic
particles \cite{books} is actually valid at all energies.
In other words although the formula (\ref{covP}) is usually derived by
taking the leading term in a $(m^2/P^2)$ series, our covariant calculation
shows that there are actually no $O(m^2/P^2)$ corrections to (\ref{covP}). 

 In
the non-relativistic regime where $P=(m_a+m_b) v /2 $, the standard kaon
oscillation formula $ cos (m_a-m_b) T $ is recovered.

There is no extra
factor
of two in the relativistic kaon oscillation 
formula  as would have appeared
without the wavepacket averaging .  We see that although
the two mass eigenstates have different proper times the wave packet
overlap results in the average proper time appearing in the interference
term. Therefore the interfernce phase is actually $(m_a-m_b) \bar{s}
=(m_a-m_b)
(m_a+m_b) T /(E_a +E_b)= (m_a^2-m_b^2) T/2E $. The average
time prescription of \cite{gold}
 can therefore be
justified using the covariant propagator method. Other methods of
showing that an extra factor of two does not appear in the flavour
oscillations formula have been discussed in \cite{lip2}.
\newpage
\section{No muon oscillation}

The energy mommentum relation was employed in the last section to express
the oscillation lenght in terms of the average mommentum of the mass
eigenstates. We can use the conservation laws to give an expicit
expression for the phase difference of the oscillation term. For example
if the neutrinos are produced by pion decay $\pi^+ \rightarrow \mu^+
\nu_{\mu}$, 
 one can use the conservation
laws to write the oscillation lenght of the neutrino in terms of $m_{\pi}$
and $m_{nu}$ and explicitly the validity of the formula for the phase
difference (\ref{cosA}). In the pion rest frame the energy and the
mommentum of the neutrino and the muon are,
\bea
E_{\nu}&=& {m_{\pi}^2 - m_{\mu}^2 + m_{\nu}^2\over 2
m_{\pi}}\nonumber\\[8pt]
E_{\mu}&=& {m_{\pi}^2 + m_{\mu}^2 -m_{\nu}^2\over 2
m_{\pi}}\nonumber\\[8pt]
P&=&|{\bf P}_{\nu}|=|{\bf P}_{\mu}| =
{((m_{\pi}^2 -(m_{\nu}+m_{\mu})^2)(m_{\pi}^2
-(m_{\nu}-m_{\mu})^2))^{1/2}\over 2
m_{\pi}}
\label{ep}
\eea
Using (\ref{ep}) we can check explicitly the expression for the neutrino
phase difference,
\bea
\Delta E_{\nu}&=& {\Delta m_{\nu}^2 \over 2 m_{\pi}} \nonumber\\[8pt]
\Delta P_{\nu}&=& {\Delta m_{\nu}^2\over 2 m_{\pi} P} (E_{\nu} -
m_{\pi})\nonumber\\[8pt]
\Delta \phi_{\nu}&=& (E_{\nu} \Delta E_{\nu} - {\bf P}_{\nu}\cdot 
\Delta {\bf P}_{\nu}){ X\over P_{\nu}}~~=~~{\Delta m_{\nu}^2 \over 2 P}X
\label{dphinu}
\eea

Another application of the conservation laws is the question of associated
particle oscillation raised by SWS \cite{SWS}. For instance in the process 
$\pi^+ \rightarrow \mu^+
\nu_{\mu}$, SWS state that since the muon mommenta depend on $m_{\nu}$,
an interfernce of muon eigenstates corresponding to different
$P_{\mu}(m_{\nu})$ will result in the occurance of oscillation in space
of the muon probability. Similarly in the reaction $p~p \rightarrow
\Lambda~K$ , SWS claim that the associated oscillations of $\Lambda$ 
probability in space ought to be seen. This claim has been disproved by
Lowe et al and Okun et al, by differnt methods. We will see that the
 phase difference of $m_{\mu}$ corresponding to the different $\nu$ mass
states actually vanishes . Using (\ref{ep}) we see that  
\bea
\Delta E_{\mu}&=& - {\Delta m_{\nu}^2 \over 2 m_{\pi}} \nonumber\\[8pt]
\Delta P_{\mu}&=& - {\Delta m_{\nu}^2\over 2 m_{\pi} P} E_{\mu} 
\nonumber\\[8pt]
\Delta \phi_{\mu}&=& (E_{\mu} \Delta E_{\mu} - {\bf P}_{\mu}\cdot 
\Delta {\bf P}_{\mu}){ X\over P_{\mu}}~~=~~0.
\label{dphimu}
\eea
Okun et al state that muon oscillations do not take place,if the neutrino
is
not
observed ' owing to the
orthogonality of the neutrino eigenstates. That would imply that if there
is a mixing of the neutrinos with some heavy states , which would make the
low-lying mass neutrinos non-orthogonal, then muon oscillations will take
place. This would have been powerful method for testing for the
orthogonality of the observed neutrinos. However from (\ref{dphimu}) we
see that
the form of the phase difference is such that the conservation laws 
ensure that whether the $\nu$ is measured or not there is no EPR
correlation to the $\mu$ phase difference. There is an EPR type 
correlation
of $E_{\mu}$ and $P_{\mu}$ with the neutrino mass eigenstates but the
phase difference vanishes . Therefore there will be no muon oscillations
In the linear combination of different mass eigenstates, niether
$\Delta E$ nor $\Delta P$ is zero. Assuming that either of them is zero
would lead to incorrect results.

\section{Conversion without oscillation}

The suppression factor $ A$ which goes with the oscillation term has
some
interesting new implications. If the parameters of the experiment are such
that $A$ becomes large then no oscillations is spacetime can be observed.
What can be observed is a constant conversion probability.

 For example,
the survival probability of a $\nu_{\mu}$  after
propagating 
over a distance $X$  in the large $A$ limit is,
( for two neutrino  flavours $(\nu_e ,\nu_{\mu})$ with the
mixing matrix
elements 
$U_{e 1} = U_{\mu 2}= cos \theta$ and $U_{e2}=-U_{\mu 1}= sin \theta$),
\be
P(\nu_\mu\rightarrow \nu_\mu;X) = sin^4 \theta+ cos^4 \theta~~
+2 sin^2\theta cos^2 \theta ~~cos({\Delta m^2 \over 2 P}X) ~e^{-A}
\simeq ~sin^4 \theta+ cos^4 \theta~~ 
\label{surv}
\ee
and the expression for the conversion probability when $A$ is large is,
\be 
P(\nu_\mu  \rightarrow \nu_{e}; t)= 2 sin^2 \theta ~cos^2\theta~
(1- cos({\Delta m^2 \over 2 P} X) ~e^{-A})~~
\simeq ~~2 sin^2 \theta ~cos^2\theta~
\label{osc}
\ee
 
It has been noted  ealier \cite{coh}
one condition $A$ must be small
which implies that $X $ must be smaller than the coherence  length
$L_{coh}=(4 {\sqrt 2} \bar \sigma P^2/\Delta m^2)$ .
      We shall show below that the
constraint $L_{osc} < L_{coh}$ is not a sufficient condition to ensure the
occurrence of flavour ocillations in space.
In the 
analysis of refs \cite{coh} only the uncertainty in the
initial position $\sigma_x$ was considered,
 in our analysis we have
included the contribution of $\sigma_t$ the uncertainty in time at which
the neutrino was produced , which as it turns out, makes the larger
contribution to the
suppression term $A$. This happens when the neutrinos are produced from
long lived resonances. The neutrino wavepacket which is produced has a
spread in time with width $\sigma_t$ which cannot be smaller than the
lifetime of the resonance whose decay produces it.
The dominant contribution in that case arises from 
the first term of $A$,
\be
A\simeq (\Delta E)^2~{\bar \sigma^2 \over 2} \simeq ({\Delta m^2
~\tau \over 2 \sqrt{ 2} E})^2
\label{A}
\ee
where we have equated the initial time uncertainity $\sigma_t$ with
$\tau$ - the
lifetime of the resonance that produces the neutrino, and we have ignored
the spatial spread of the wave-packet $\sigma_x$ which in most
experiments is many  orders of magnitude smaller than $\sigma_t$. 
In terms of the oscillation length $ L_{osc}= 4 \pi E/ (\Delta m^2)$
the suppression factor $A$. 
In order to observe oscillations the  source-detector distance
$L$ must be larger than $L_{osc}$.
Which  that the
minimum source-detector distance in order that neutrino oscillations be
observed is given by condition,
 $~L_{min}~=~(\pi)\sqrt{2} \tau $.
We shall show below that most neutrino experiments do not satisfy this
criterion for observability of space oscillations. On the other hand
spatial oscillations of flavour  can be observed in the $\Phi$  and $B$
factories owing to the large width of $\phi$ and $\Upsilon$ resonances.

For relativistic particles produced from long lived resonances the 
formula for the conversion probability 
which should be fitted with the experiments is ,
\be
P(\nu_{\alpha} \rightarrow \nu_{\beta})={1\over2}~sin^2 2\theta ~
~~\left(1-~cos({2.53 \Delta m^2 L \over E})~~exp\{-({1.79 \Delta m^2 \tau
\over
E})^2\} ~~\right)
\label{cor}
\ee 
where $\Delta m^2$ is the mass square difference in $eV^2$ ,$L$ is the
detector distance in $m~(km)$,  $\tau$ is the lifetime of the parent
particle
in the lab frame
in $m~(km)$ and $E$ is the  energy
in $MeV~(GeV)$.

\section{Constraints on $\Delta m^2$ and $sin^2 2 \theta$ from
experiments}

 We shall now fit the oscillation formula (\ref{cor}) with 
  the results of some terrestrial
 neutrino oscillation experiments ,                           
  classified according to the type of the source
 of neutrinos.

%\item
 {\it Stopped Muons    :}
The LSND experiment \cite{LSND} has searched for $\bar \nu_{\mu}
\rightarrow \bar \nu_e$ in a beam of anti-muon neutrinos produced from
stopped muons. The minimum uncertainty in the time of $\bar \nu_{\mu}$
production is the muon lifetime $ \tau_{\mu}= 658.65 m$.
The average neutrino energy $E_{\nu} \simeq 30 MeV$ and the detector is
at
a distance $L= 30 m$ from the source.
In order that neutrino oscillations from muon decay neutrinos be seen
the minimum source detector distance must be $L_{min}=\sqrt {2} \pi
\tau_{\mu}= 2927.6 m$.
This means that the conversion probability does not oscillate as a 
function
of distance. Fitting the reported \cite{LSND}
range of conversion probability  $P(\bar
\nu_{\mu} \rightarrow \bar \nu_e)= (0.31 + 0.11 -0.1 \pm 0.05)\times
10^{-2}$ with the formula (\ref{cor}) ,
the regions of allowed $\Delta m^2$ and $sin^2 2\theta$ which gives rise
to this probability band are shown as dotted band in Fig. 1. The same
range of conversion probability when fitted with the standard oscillation 
formula gives the shaded region shown in Fig.1.
We have averaged the conversion probability  by assuming a gaussian 
distribution of neutrino energy, with average energy $30 MeV$ and a half
width  of $ 10 MeV$. 
The asymptotic bounds are as follows. If $
\Delta m^2 > (~~2\sqrt{2} E / \tau_{\mu} )= 2.54 \times 10^{-2}
eV^2$
then  mixing angle  constrained
in the range
$0.3 \times 10^{-2} \leq  sin^2 2\theta \leq 0.9 \times 10^{-2} $.
For large mixing angles $\sin^2 2\theta \simeq 1 $ the mass square
difference lies in the range, 
$2\times 10^{-3} eV^2 \leq \Delta m^2 \leq 3 \times 10^{-3}
eV^2 $.
This bound is two orders lower than what is obtained by the use the
standard oscillation formula \cite{LSND}.

%\begin{itemize} 

%\item
 {\it Stopped Pions :}
The Karmen experiment \cite{karm} searches for 
$\nu_{\mu} \rightarrow \nu_e$ conversion in a beam of $\nu_{\mu}$ produced
by the decay of
stopped pions . The time at which the $\nu_{\mu}$ is prduced cannot be
localised to less than the pion lifetime $ \tau_{\pi^+}=7.804 m$. The
$\nu_{\mu}$ energy is $E_{\nu}= 29.8 MeV$. In order for
oscillations to be observable the detector distance must be larger than
$L= \sqrt 2 \pi \tau_{\pi^+}=35 m$. Since the detector 
 distance is  $17.5 m$, it is not possible to observe
spatial oscillations in the Karmen experiment. 
The experimental bound  $P(\nu_{\mu}\rightarrow \nu_e) < 3
\times10^{-3}$,
fitted with the  conversion probabilty formula (\ref{cor}) constrains
 $sin^2  2\theta < 0.6 \times 10^{-2}$ if $\Delta m^2 > (2 \sqrt {2} E
/\tau_{\pi^+})= 2 ~~eV^2 $.  In the $sin^2 2\theta \sim 1$ limit 
$\Delta m^2  < 0.16 eV^2  $. 
We have averaged the conversion probability  by assuming a gaussian 
distribution of neutrino energy, with average energy $30 MeV$ and a half
width  of $ 10 MeV$. 
The regions
of the parameter space allowed by the covariant wavepacket formula and by
the standard formula are shown in Fig.2. Since the pion decay length is of
the same order as the experimental baseline, the improvement in sensitivity
to $\Delta m^2$ is marginal compared to LSND.

%\item 
{\it High energy Pions or Kaons:}
When high energy neutrino beams are obtained from the decay of a pion or 
kaon in
flight the both $\tau$ and $E$ are larger by the Lorentz factor $\gamma$
and consequently $A$ remains the same as that with stopped pions or kaons.
In the BNL E776 experiment \cite{E776} most of the neutrinos are
produced in
$\pi^{\pm}$ decay and have an average energy $E\simeq 5 Gev$ and the
detector is at a distance $L=1 km$ . 
In the lab frame the neutrino time uncertainty is $\tau=
\tau_{\pi^+} (E_{\pi^+} /m_{\pi^+})$. Taking the
mean $\pi^{\pm}$ energy to be $10 GeV$ , the lifetime of the
pions in the lab frame is $\tau= 7.804 ( 10/0.135)m = 0.578 km$. Fitting
the E776 experimental limit $P(\nu_{\mu} \rightarrow \nu_e) < 1.5 \times
10^{-3} $ with the conversion probability formula (\ref{cor}),
we see from Fig.3 that 
$sin^2 2\theta < 0.3 \times 10^{-2}$ if $\Delta m^2 > (2\sqrt{2} E
/\tau_{\pi^+})= 5 ~~eV^2 $.  In the $sin^2 2\theta \sim 1 $ limit ,
 $\Delta m^2  < 0.2 eV^2  $ .

 In the CCFR \cite{CCFR} experiments most of the neutrinos are produced in
$K^{\pm}$ decay and have an average energy $E\simeq 140 Gev$ and the
detector is at a distance $L=1.4 km$ . 
In the lab frame the neutrino time uncertainty is $\tau=
\tau_{K^+} (E_{K^+} /m_{K^+})$. Taking the
mean $K^{\pm}$ energy to be $600 GeV$ we see that the lifetime of the
kaons in the lab frame is $\tau= 3.7 ( 600/0.493)m = 4.514 km$. Fitting
the CCFR experimental limit $P(\nu_\mu \rightarrow \nu_e) < 0.9 \times
10^{-3} $ with the conversion probability formula (\ref{cor}) we see from
Fig.3 that
$sin^2  2\theta < 1.8 \times 10^{-3}$ if $\Delta m^2 > (2 \sqrt{2} E
/\tau_{K^+})= 76 ~~eV^2 $.  In the $sin^2 2\theta \sim 1$ limit ,
 $\Delta m^2  < 1.2 eV^2  $.

%\item 
{\it Reactor neutrinos:}
Reactor neutrinos are produced primarily from the beta decay of $^{235}U$
and $^{229} Pu$ which have a half life $\sim 10^2 sec$ \cite{BUGEY}.
 The minimum detector
distance for observing neutrino oscillations is $10^{11} m$
 which means
that it is impossible to detect spatial oscillations with reactor
neutrinos. The upper bounds on conversion probability in the
Bugey experiment is $P(\bar \nu_e \rightarrow \bar \nu_{\mu}~ or~
\bar\nu_{\tau}) < 0.075 $.  The largest detector distance $L=95 m$ and the
average neutrino energy is $E\simeq 5 MeV$. 
The conversion probability formula is independent
of $\Delta m^2 $ if $\Delta m^2 > 10^{-9} eV^2$. Fig 3. shows that in
this limit the reactor experiments rule out   $sin^2 2\theta \geq 0.15$.
In the standard oscillation formula the oscillatory term averages to zero 
 when $\Delta m^2 >(4\pi E/L)$. Using the
wavepacket
formula (\ref{cor}) however we see that the oscillatory term is 
is exponentially damped at much lower values of $\Delta m^2$  .
If neutrino sources have a large decay time  which is
the case in most experiments , the conversion probabilty is sensitive to
mass differences  $\Delta m^2 > (2 \sqrt 2 E/\tau)$. 
For this reason the LSND
experiment which uses neutrinos from muon decay has two orders of
magnitude more sensitive than  KARMEN, BNL-E776 etc  where the neutrinos
are from $\pi$ and $K$ decay. 
This is evident from the combined plot of these experimental results, 
shown in Fig.3 fitted
with the covariant wavepacket formula (\ref{cor}).
The reactor experiments where $\tau \sim
10^2 s$ are in fact
sensitive to values of $\Delta m^2 $ as low as $10^{-9} eV^2$. But the
probability measurement in reactor experiments is poor compared to the
accelerator experiments
which is why they rule out only a small region of parameter space.  

%\item
{\it  Meson oscillations at $B$ and $\Phi$ factories:} The 
mesons produced by $\phi$ or $\Upsilon(4S)$ decay are non-relativistic
and the expression for the damping factor is $A=2 (\Delta m ~\tau)^2$. For
the kaons produced from from $\phi$ decays, $\Delta m_K= 3.5 \times
10^{-12}
MeV $ and the $\phi$ lifetime is $\tau_{\phi}= 0.225 MeV^{-1}$, and the
damping factor $A= 2.4 \times 10^{-24}$
  is  negligible. 
For the case of $B \bar B $ oscillations
produced from
$\Upsilon(4S)$ decays, the $\Delta m_B =3.53 \times 10^{-10} ~MeV$ and 
the lifetime of $\Upsilon(4S)$ is $\tau_{\Upsilon}=0.042 MeV^{-1}$ and
again the damping factor, $A=8.6 \times 10^{-22}$ is negligible. This is
important
for tests of EPR \cite{EPR} and  Quantum Mechanics \cite{QM} 
where the measurement of the oscillatory
term of the probability is essential.
 %\end{itemize}

\section{Conclusions} 
The covariant formulation gives two new results compared to the standard 
treatment. We use the condition that the spread of the neutrino wave
function along the time axis is given by the lifetime of the particle
whose decay produces the neutrino. When neutrinos are produced from long
lived particles like muons , the interference term in the conversion
probability formula vanishes for much smaller values of the neutrino
mass square difference compared with the standard formula. The reason
this happens can be understood with the following picture. When the spread
of the neutrio is large along the time axis, its spread along the energy
axis is small. The neutrino wavefunctions for different masses are then
energy eigenstates which are mutually orthogonal. The interference term
which is the overlap of the wavefunctions of diffferent mass states
therefore vanishes when these states become orthogonal.  
 
The second new result is that the for the secondary particles, like
muons in the decay $\pi \rightarrow \mu \nu$ there is no spacetime
variation of probability independent of whether the neutrinos are
measured or not. Although the muon wavefunction
is a linear combination of distinct energy and momentum states
corressponding to
the different $\nu$ mass states, the phase difference for the
secondary particles vanishes. Therefore there is no EPR type oscillation
of the secondary particles probability distribution. This is true
independent of whether the primary oscillations of the $\nu$'s  
is measured or not. This conclusion is results from the application
of the correct conservation laws according to which the different mass
states of $\nu$  have neither energy nor momentum in common
as is sometime assumed in the standard derivations.

{\it Acknowledgments} I thank Terry Goldman, Walter Grimus and Marek
Nowakowski for useful correspondence.

.
\begin{figure}
\vskip 20cm
\includegraphics{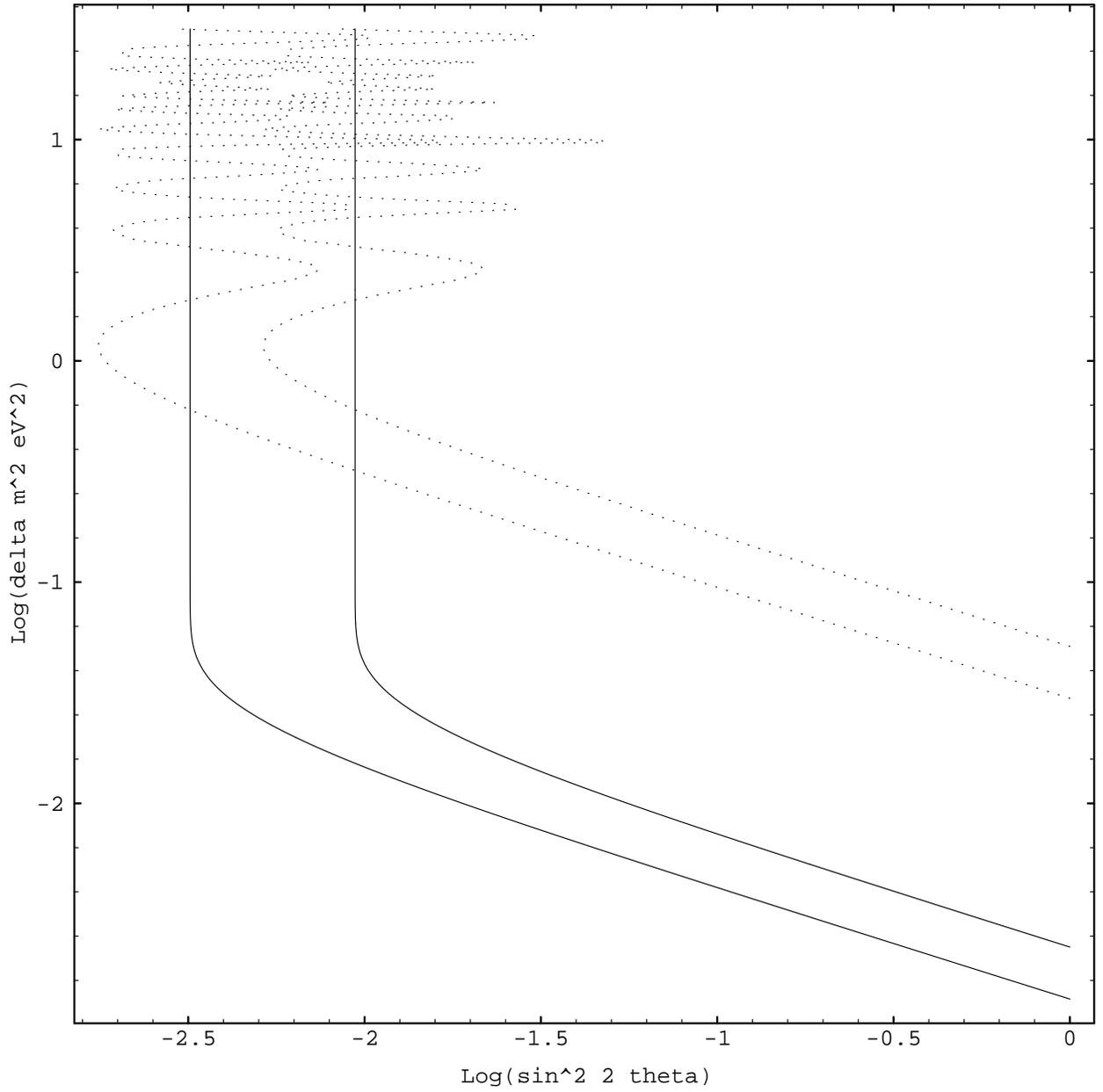}
\caption[dummy]{ Lsnd $\mu$ decay at rest experiment \cite{LSND} allowed
regions with
the standard formula
(between the dotted
curves)and the covariant
oscillation formula (continuous curves). }

\label{fg:lsnd}
\end{figure}
\vskip 0cm

\begin{figure}
\vskip 20cm
\includegraphics{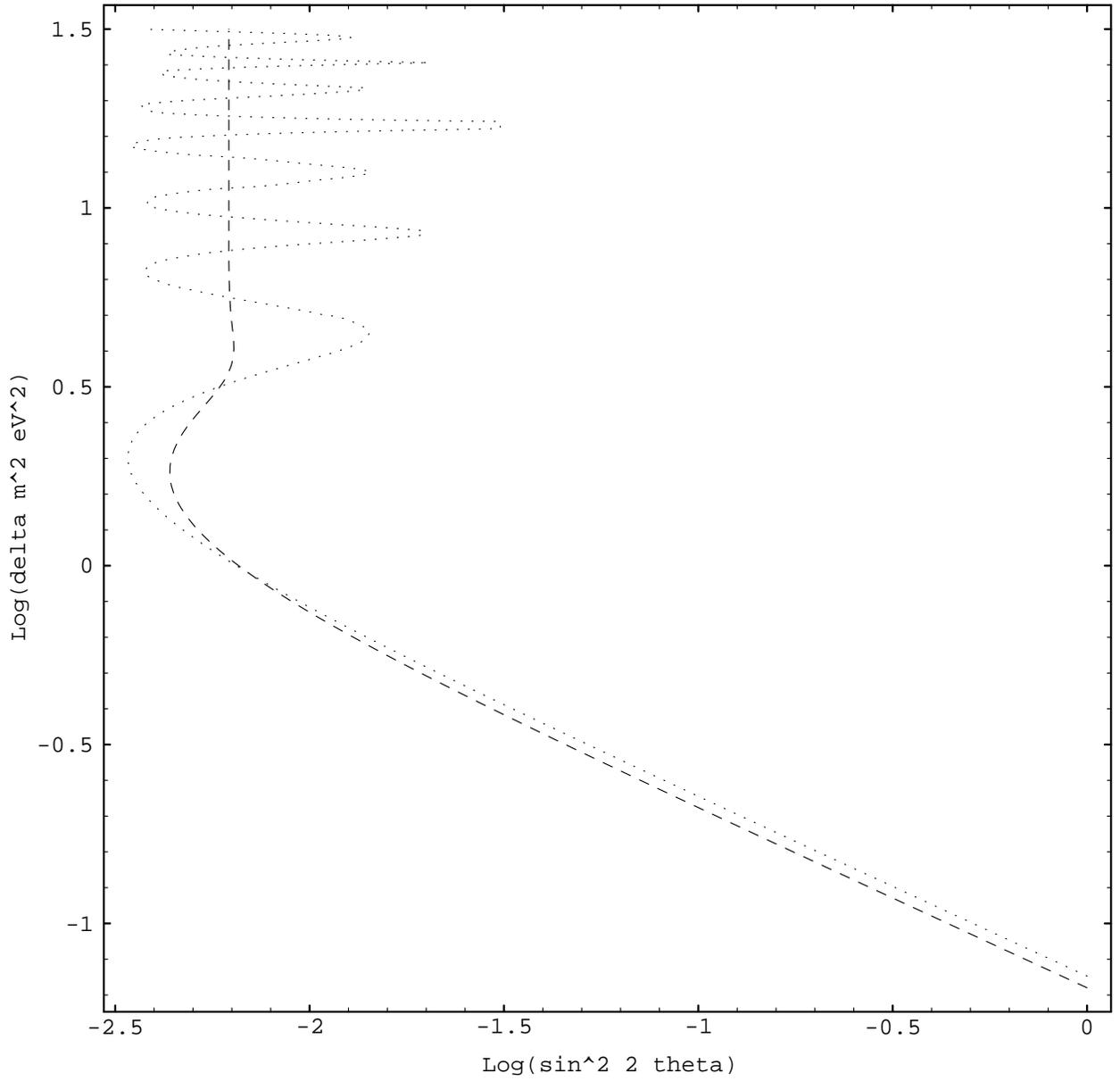}
\caption[dummy]{ Karmen $\pi$ decay at rest experiment \cite{karm} allowed
regions with the standard formula
(dotted line) and the
covariant oscillation formulas (dashed line).}
\label{fg:karm}
\end{figure}
\vskip 0cm

\begin{figure}
\vskip 20cm
\includegraphics{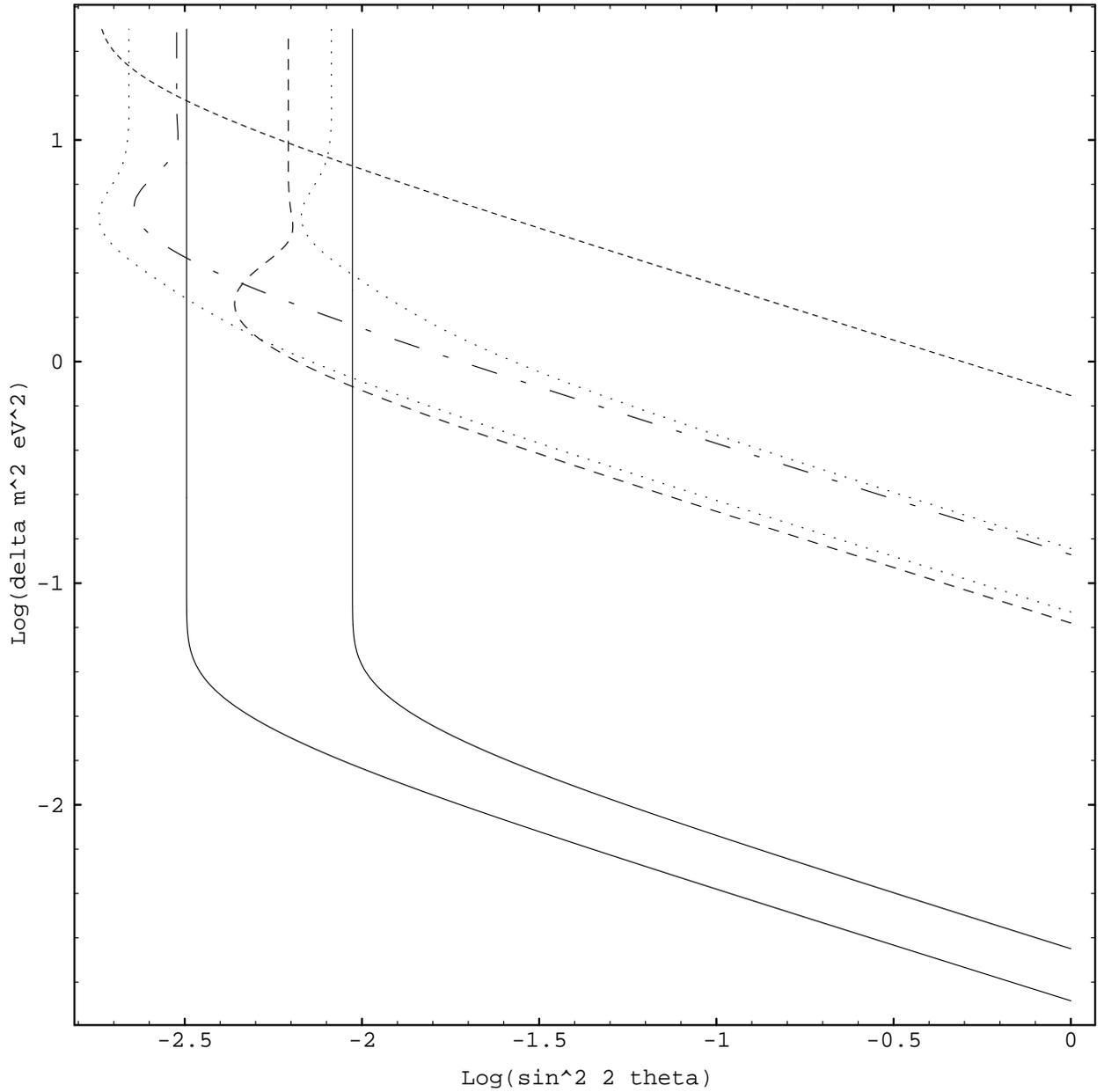}
\caption[dummy]{The   region between the continuous lines is allowed by
the LSND $\mu$
decay in flight  experiment \cite{LSND} .The region between the dotted
lines is
allowed by the
LSND $\pi$ decay in flight experiment \cite{LSNDpi} . Region ruled out
by E776 \cite{E776} is above the dashed-dotted curve and by Karmen
\cite{karm} is above dashed
curve. Region above the topmost dashed curve is ruled out by CCFR
\cite{CCFR}.}
% \label{fg:all}
\end{figure}

\end{document}